\documentstyle[12pt]{article}
\begin{document}

\centerline{\Large Work with Apple's Rhapsody Operating System}
\vskip 4pt
\centerline{\large which Allows Simultaneous UNIX Program Development,}
\vskip 4pt
\centerline{\large UNIX Program Execution, and PC Application Execution}
\bigskip
\centerline{Don Summers, Chris Riley, Lucien Cremaldi, and David Sanders}
\vskip 1pt
\centerline{Dept.~of Physics, Univ.~of Mississippi--Oxford, University, MS
38677}
\vskip 1pt
\centerline{summers@umsphy.phy.olemiss.edu} 
\bigskip
\bigskip
\leftline{\bf ABSTRACT}

   Over the past decade, UNIX workstations have provided a
very powerful program development environment.  However,
workstations are more expensive than PCs and Macintoshes
and require a system manager for day-to-day tasks such
as disk backup, adding users, and setting up print queues.
Native commercial software for system maintenance and ``PC
applications'' has been lacking under UNIX.

   Apple's new Rhapsody operating system puts the current Mac\,OS
on a NeXT UNIX foundation and adds an enhanced NeXTSTEP object oriented 
development 
environment called Yellow Box.
Rhapsody simultaneously runs UNIX and commercial Macintosh
applications such as word processing or spreadsheets.  Thus a UNIX detector
Monte Carlo can run for days in the background at the same
time as a commercial word processing program. And commercial
programs such as Dantz Retrospect are being made available to
make disk backup easy under Rhapsody.

   Apple has announced that in 1999 they intend to
be running Rhapsody, or Mac\,OS X as it will be called in the commercial 
release,
on all their newer computers.
Mac\,OS X may be of interest to those who have
trouble hiring expert UNIX system managers; and to those who
would prefer to have a single computer and operating system
on their desktop that serves both the needs of UNIX program
development and running commercial applications, simultaneously.

   We present our experiences running
UNIX programs and Macintosh applications under the Rhapsody DR2 
Developer Release.

\vfill 
\centerline{CHEP'98  \ Computing in High Energy Physics Conference}
\vskip 1pt
\centerline{Commodity Hardware \& Software Session}
\vskip 1pt
\centerline{August 31 -- September 4, 1998}  
\vskip 1pt
\centerline{Hotel Inter-Continental, \ Chicago, Illinois}
\vskip 1pt
\centerline{http://www.hep.net/chep98/index.html}
\vskip 1pt
\centerline{Abstract \#197}
\vfill
\eject
\leftline{\bf HISTORICAL INTRODUCTION}
\medskip

   Historically, UNIX Workstations from Sun, IBM, HP, DEC, and SGI have
provided powerful platforms and farms for high energy physics [1,\,2]. UNIX is
a versatile development environment and UNIX Workstations come with excellent,
well optimized C, C++, and Fortran high level language compilers. However,
workstations now cost more than PCs and Macintoshes with comparable power. And
workstations are weak in ``PC applications'' such as word processors
or spreadsheets, so
many people put two computers on their desktop.

   Workstations are also 
lacking in {\it point and click} system management tools
(e.g. Dantz Retrospect [3] for incremental disk backups over a network).
UNIX workstations usually require finding and paying  a system manager for 
day-to-day maintenance.

\bigskip
\bigskip
\leftline{\bf LINUX}
\medskip

The Linux operating system runs on inexpensive PCs and Macintoshes. The Red
Hat Linux CD only costs \$50 for x86 computers with CPUs such as the Pentium
II and AMD K6--2.  MkLinux [4] for the PowerPC CPU is available from Apple and
runs on top of a Mach kernel.  LinuxPPC [5] is a monolithic port of Linux to 
the PowerPC and runs native rather than on top of Mach.
Linux has the gnu C, C++, and Fortran compilers.
Linux provides the versatile development environment one expects from UNIX.

However, Linux is  weak in ``PC office suites,'' so
many people dual boot Linux and Windows. (Some applications like 
Corel WordPerfect 8 have recently become available. And work is in
progress on other applications such as the Objectivity database,
Lotus SmartSuite, and Corel's complete business application suite.) 
If one was running both Linux and Windows NT alternately
on a PC, a long event reconstruction run would have to be
stopped just to run an Excel spread sheet. With two operating
systems running alternately, a system manager would have to
separately catch each running operating system to backup all disks. 

One
could run a Windows NT only environment, but Windows NT is
primarily an applications environment. Creating analysis
software is much easier under UNIX, which is more robust and has been designed
for program development.

\vfill
\eject
\leftline{\bf APPLE RHAPSODY DR2 and Mac\,OS X}
\medskip

   Apple is combining Mac\,OS with NeXT UNIX for their PowerPC based
computers. By next year, Apple plans to have all of their newer Macintoshes
running Mac\,OS applications and BSD 4.4 UNIX on a Mach 3.0 kernel
foundation, which will allow symmetric multi-processing, good multi-tasking,
and memory protection.  An enhanced NeXTSTEP/OpenSTEP object oriented 
development environment is supplied and called Yellow Box. 

The developer's release is called Rhapsody DR2 and the
commercial release next year is to be called Mac\,OS X [6].  The X in 
Mac\,OS X is the Roman numeral representation of the number 10.

  The NeXT UNIX development environment [7] was used by Tim Berners-Lee at CERN
to invent the World Wide Web in 1989.
The actual NeXT computer that ran the original World Wide Web server and
browser at CERN may be seen in Physics Today [8].

Compilers which generate well optimized code and don't introduce bugs are a
key component of any operating system.
The Rhapsody DR2 beta release comes with the gnu C and C++ compilers and the
NeXT objective C compiler.
MetroWerks' C, C++, and Java compilers [9] are under development for Mac\,OS X
as are the Absoft
C, C++, Fortran 77, and Fortran 90 compilers [10]. 
The current Mac\,OS 8 
Absoft Fortran compiler is being used for plasma physics simulations
on Macintoshes with PowerPC G3 CPUs at the University of 
California--Los Angeles [11].
Motorola has a library of routines for speeding up floating 
point operations on the PowerPC [12].
MetroWerks' CodeWarrior Pro 4 supports Motorola's
forthcoming
AltiVec vector processing unit [13].  
IBM AIX RS/6000 compilers [14] run on
PowerPC, but so far no ports to Mac\,OS X are publicly in progress.  
The IBM compilers
appear to be very good. 

  Applications including word processors, spreadsheets, and web browsers run on
the Rhapsody
DR2 beta release. 
An ssh (Secure Shell) package is available [15].
Dantz already has a beta version of their Retrospect disk backup software
for Rhapsody.  
When Mac\,OS X arrives with its new {\it Carbon} application
interface library, applications which add the modest modifications required by 
{\it Carbon} will gain the advantages of UNIX such as multi-tasking and memory
protection.  The {\it Carbon} Application Interface library updates one-fourth
of the 8\,000 Mac\,OS 8 APIs. Applications based on the old Mac\,OS 8 APIs will 
still
run via what Apple calls the Blue Box.  
While a browser crash in Blue Box may 
take the whole Blue Box down, 
it WILL NOT
take down the whole machine.  The Yellow Box APIs also take advantage of
multi-tasking and memory protection and provide an object oriented
development environment.  Yellow Box applications can also run under 
Windows 98 and Windows NT.

  Dual booting or having to have two different computers on your desk may 
become history.
One may now be able 
run number crunching programs in the background, at low priority, 
at the same time as
a secretary runs word processing.  We have had good experience running
event reconstruction in the background on {\it desktop} workstations [2].
The goal is to do much more 
with the computers one can afford to buy.
Setting up print queues, adding users, and other system
tasks may be simplified.  The goal is not to have to devote a UNIX guru
to day-to-day maintenance.

\vfill
\begin{table}[h]
\begin{center}
\renewcommand{\arraystretch}{1.15}
\begin{tabular}{lc} \hline \hline
Operating System Release &   Estimated Date  \\ \hline
Rhapsody DR2 Developer Release      &  May 1998 \\
Mac\,OS X Server Commercial Release  & January 1999 \\
Mac\,OS X beta with {\it Carbon} Libraries &  Q1 1999 \\
Mac\,OS X with {\it Carbon} Libraries      &  Q3 1999 \\ \hline \hline
\end{tabular}
\caption{Apple Operating System Upgrade Timeline.}
\end{center}
\end{table}
\begin{table}[h]
\begin{center}
\renewcommand{\arraystretch}{1.15}
\begin{tabular}{|c|c|c|c|} \hline 
Blue Box     & {\it Carbon}   & Yellow Box   & BSD UNIX 4.4    \\ 
Mac APIs     & Updated Mac APIs & NeXT OO APIs  & Applications    \\ \hline
\multicolumn{4}{|c|}{Mach 3.0 Kernel} \\ \hline 
\end{tabular}
\caption{Mac\,OS X Architecture. Old Mac\,OS applications run in the Blue
Box without any changes.  If one Blue Box application crashes all 
applications in the Blue Box may crash. One--fourth of the 8\,000 Mac\,OS
APIs were updated for {\it Carbon} to add advantages like memory protection.
Greg Gilley, vice president of graphics products at Adobe Systems, ported 
Photoshop 5.0 to {\it Carbon} in nine days.   The Yellow Box provides an 
enhanced NeXTSTEP/OpenSTEP object oriented 
development environment. Blue Box, {\it Carbon,} Yellow Box, and BSD UNIX 4.4
all run on top of the Mach 3.0 kernel.}

\end{center}
\end{table}
\eject
\leftline{\bf EXPERIENCE WITH RHAPSODY DR2}
\medskip

We have compiled and run simple C programs.
ftp, telnet, rlogin, and multiple user logins work. 
Software is included for mail servers and DNS name servers.
Mac applications do run in Apple's Blue Box with no apparent slowdown.

We have downloaded the gnu g77 Fortran compiler [16] for a PowerPC Macintosh
running MkLinux and have run moderately long Fortran programs. 

With Apple's assistance, 
we are working to port the gnu g77 Fortran compiler [17] to Rhapsody.
We plan to run the Fermilab E769 benchmark program.
This is a 60\,000 line Fortran program that does track and calorimeter 
reconstruction [1].

We are considering a port of CERN's GEANT Monte Carlo program to this new 
flavor of UNIX.

\bigskip
\bigskip
\leftline{\bf PowerPC CPUs}
\medskip
Current Macintoshes run using the PowerPC 750 CPU at speeds up to 333 MHz. The
CPU is also called the PowerPC G3 and is made by IBM [18] and 
Motorola [19]. 

The PowerPC G4 [20] is under development by IBM and Motorola for 1999.  
As compared to the G3, 
it doubles the width of the bus to the back side cache to 128 bits 
and doubles the maximum cache size to 2MB. It adds the
option of doubling the bus width to main memory to 128 bits.
The G3 has a 32-bit multiplier; floating point multiplies 
require two cycles.
The G4 has a full 64 bit multiplier to improve floating point performance.
The PowerPC G4 adds support for multiple processors so that more than one
CPU can use the same cache without having to move data back to main memory.
There is talk of putting four G4 processors and their cache on the same piece
of silicon.  The low power consumption of the PowerPC architecture allows this
and also makes the PowerPC G4 an ideal candidate for even faster operation at
low temperature ($-40^0$C [21]).  The move from aluminum to copper chip
interconnects will decrease $I^2R$ power losses, decrease $RC$ time constants,
and raise heat conductivity
as does operation at low temperature.
The resistivity of copper is 40\% less than aluminum.

Motorola intends to add the AltiVec 128-bit SIMD processor unit 
[13] to its PowerPC 
G4 processors.  
The acronym SIMD stands for Single Instruction, Multiple Data.
The AltiVec processing unit 
includes 32 128-bit registers and 162 new instructions, mostly
for multimedia.
However, new instructions do include the ability to do four 32-bit floating 
point multiplies 
and four
32-bit floating point adds or to permute 128-bits by any pattern in a single 
clock cycle. The AltiVec can use the
Newton--Raphson algorithm for division and square roots.

Low power, embedded PowerPC chips include the 200 MHz IBM PowerPC 405 [22]
with a 16KB data cache, a 16KB instruction cache, and an Integer Multiply and
Accumulate unit allowing Soft Modem support. Code compression, Universal 
Serial Bus, Firewire,
Accelerated Graphics Port, ethernet, IrDA, VGA, and Rambus memory control are
some of the options [23] available on the same chip with the PowerPC core. IBM
also has a license for AltiVec.

\begin{table}[h]
\begin{center}
\renewcommand{\arraystretch}{1.15}
\begin{tabular}{lc} \hline \hline
CPU &   Gigaflops  \\ \hline
Motorola PowerPC G4 AltiVec  &  3.0 \\
Cray SV1                  &  4.0 \\
AMD K6--2 3Dnow! (350MHz) &  1.4 \\
Intel Pentium II (450MHz)       &  0.45 \\ \hline \hline
\end{tabular}
\caption{Gigaflop Ratings of CPUs.
Integer performance of these CPUs does not vary as widely as the floating
point performance.
MetroWerks' CodeWarrior Pro 4 C and C++ compilers [9] provide support for the
AltiVec and 3DNow! vector processing instruction sets.}
\end{center}
\end{table}

\bigskip
\bigskip
\leftline {\bf SUMMARY -- DOING MORE WITH LESS}
\medskip

Apple is moving to BSD 4.4 UNIX next year as a part of Mac\,OS X.
Most Macintosh users will just see the UNIX advantages of memory protection
and multitasking.  One application can no longer crash a whole computer.
The UNIX command line will not be at all necessary to run applications.

UNIX users may be able to run both ``PC applications'' and develop and
run UNIX jobs on one computer with one operating system, Mac\,OS X.  
Spreadsheets and word processing can run at the same time as number crunching
jobs.
More work can
be done on fewer computers.

Automated disk backup and other {\it point and click} administration tools are
available for Mac\,OS.  UNIX gurus may be freed to
write programs rather
than just maintain a system.

\def\issue(#1,#2,#3){\space{\bf{#1}}\space(#2)\space#3}

\end{document}